\documentstyle[11pt,epsfig,here]{article}

\def\m0{\hbox{$m_{\hbox{\scriptsize 0}}$}}

\def\A0{\hbox{$A_{\hbox{\scriptsize 0}}$}}

\begin{document}
\vspace*{-1.8cm}
\begin{flushright}
\flushright{\bf LAL/RT 00-01}\\
\vspace*{-0.2cm}
\flushright{February 2000}
\end{flushright}
\vskip 2.5 cm
%
%
%
\begin{center}
{\bf ANALYTICAL TREATMENT OF THE EMITTANCE 
GROWTH\\
IN THE MAIN LINACS OF FUTURE LINEAR COLLIDERS}
\end{center}
\vspace {1.2 cm}

\begin{center}
{ \large \bf Jie Gao}   
\end{center}     

\begin{center}
{\Large \bf Laboratoire de l'Acc\'el\'erateur Lin\'eaire\\}
 {IN2P3-CNRS et Universit\'e de Paris-Sud, BP 34, F-91898 Orsay Cedex}
\end{center}     

\vspace {1 cm}
\begin{abstract}
In this paper the single and multibunch emittance growths 
in the main linac
of a linear collider are analytically treated in analogy to the
Brownian motion of a molecule, and the analytical
formulae for the emittance growth due to accelerating structure
misalignment errors are obtained by solving Langevin equation.
The same set of formulae is derived also
by solving directly Fokker-Planck equation. Comparisons with numerical
simulation results are made and the agreement is quite well. 
\end{abstract}
\vspace {0.5cm}
\section{Introduction}
To achieve the required luminosity in a future e$^+$e$^-$
linear collider one has to produce two colliding beams at the 
interaction point (IP) with extremely small transverse beam dimensions. 
According to the linear collider design principles described in ref. 1,
the normalized beam emittance in the vertical plane (the normalized
beam emittance in the horizontal plane is larger) 
at IP can be expressed as:    
\begin{equation}
\gamma \epsilon_y={n_{\gamma}^4r_e\over 374\delta_B^*\alpha^4}
\label{eq:1a}
\end{equation}
where $\gamma$ is the normalized beam energy,
$r_e=2.82\times 10^{-15}$ m is the classical electron radius,
$\alpha=1/137$ is the fine structure constant,
$\delta_B^*$ is the maximum tolerable beamstrahlung energy spread,
and $n_{\gamma}$ is the mean number of beamstrahlung photons per electron
at IP. Taking \linebreak $\delta_B^*=0.03$ and $n_{\gamma}=1$, one finds
$\gamma \epsilon_y=8.86\times 10^{-8}$mrad. To produce beams of this small
transverse emittance damping rings seem to be the only known facilities
which have the potential to do this job. The questions now are that 
once a beam of this
small emittance is produced at the exit of the damping ring, how about the 
emittance growth during the long passage through the accelerating
structures and the focusing channels from the damping ring at the
beam energy of few GeV to the IP with the beam energy of a few hundred of GeV,
and how to preserve it ? 
\newpage
Many works have been dedicated to answer these questions
\cite{Chao1}\cite{Balakin}\cite{Tor}\cite{PYJ}. 
To start with, in sections 2, 3, and 4, we consider the 
short range wakefield induced single bunch emittance growth and 
try to calculate the emittance growth
by using two different methods, and show that the two methods give the same
results. 
Since the number of accelerating structures in the main linac of a linear 
collider is very large, the transverse random kicks on the particles can be
described statistically. 
Firstly, we make use of the analogy between the transverse motion of
a particle in a linear accelerator with the Brownian motion of a molecule,
which are governed by Langevin equation.
Secondly, we solve directly Fokker-Planck equation. What should be noted is that
both methods are physically consistent.
As a natural extension, multibunch case is treated in section 5. Comparisons with some
numerical simulation results are made in section 6.  
\par
\section{Equation of transverse motion}
The differential equation of the transverse motion of a bunch with zero 
transverse dimension is given as: 
$${d^2y(s,z)\over ds^2}+{1\over \gamma (s,z)}{d\gamma (s,z)\over ds}{dy(s,z)
\over ds}
+k(s,z)^2y(s,z)
$$
\begin{equation}
={1\over m_0c^2\gamma (s)}
e^2N_e\int_z^{\infty}\rho (z'){\cal W}_{\perp}(s,z'-z)y(s,z')dz'
\label{eq:1}
\end{equation}
where $k(s,z)$ is the instantaneous betatron wave number at position $s$, 
$z$ denotes the particle
longitudinal position inside the bunch, and 
$\int_{-\infty}^{\infty}\rho (z')dz'=1$.
Now we rewrite eq. \ref{eq:1} as follows:
\begin{equation}
{d^2y(s,z)\over ds^2}+\Gamma{dy(s,z)\over ds}
+k(s,z)^2y(s,z)=\Lambda
\label{eq:2}
\end{equation}
where $\Gamma ={\gamma(0)G\over \gamma (s,z)}$, 
$G={eE_z\over m_0c^2\gamma (0)}$,
$E_z$ is the effective accelerating gradient in the linac, \linebreak 
$\Lambda ={e^2N_eW_{\perp}(s,z){y(s,0)}
\over m_0c^2\gamma (s,z)}$, 
$W_{\perp}(s,z)=\int_z^{\infty}\rho (z'){\cal W}_{\perp}(s,z'-z)dz'$,
and ${y(s,0)}$ is the deviation of the bunch head with
respect to accelerating structures center. 
In this section we consider the case
where the injected bunch, 
quadrupoles and beam position monitors are well aligned, while
the accelerating structures are misaligned. As a consequence,
$y(s,0)$ is a random variable exactly the same as
random accelerating structure misalignment errors
with $<y(s,0)>=0$ ($<$ $>$ denotes the average over $s$).
If we take $z$ as a parameter and regard $\Gamma$, $k(s,z)$, and $\Lambda$ 
as adiabatical variables with respect to $s$, eq. \ref{eq:2} can be regarded
as Langevin equation which governs the Brownian motion of a molecule.
\par
\section{Method one: Langevin equation}
To make an analogy between the movement of
the transverse motion of an electron and that of a molecule, 
we define $P ={e^2N_eW_{\perp}(s,z)l_s
\over m_0c^2\gamma (s,z)}$, and regard $y(s,0)P$ as the particle's "velocity"
random increment ($\Delta {dy\over ds}$) 
over the distance $l_s$, where $l_s$ is 
the accelerating structure length. 
What we are interested is to 
assume that the random
accelerating structure misalignment error follows Gaussian distribution: 
\begin{equation}
f(y(s,0))={1\over \sqrt{2\pi}\sigma_y}\exp\left(-{y(s,0)^2\over 2\sigma_y^2
}\right)
\label{eq:7a}
\end{equation}
and the velocity ($u$) 
distribution of the molecule follows Maxwellian distribution:
\begin{equation}
g(u)=\sqrt{{m\over 2\pi kT}}\exp\left(-{mu^2\over 2kT
}\right)
\label{eq:7aa}
\end{equation}
where $m$ is the molecule's mass, $k$ is the Boltzmann constant, and $T$
is the absolute temperature.
The fact that the molecule's velocity follows
Maxwellian distribution permits us to get the distribution function
for $\Lambda l_s$ \cite{Chand}:
\begin{equation}
\phi(\Lambda l_s)={1\over \sqrt{4\pi ql_s}}\exp\left(-{\Lambda^2
l_s^2\over 4ql_s}\right)
\label{eq:7b}
\end{equation}
where
\begin{equation}
q=\Gamma {kT\over m}
\label{eq:7d}
\end{equation}
By comparing eq. \ref{eq:7b} with eq. \ref{eq:7a},
one gets:
\begin{equation}
2\sigma_y^2={4ql_s\over P^2}
\label{eq:7c}
\end{equation}
or
\begin{equation}
{kT\over m}={\sigma_y^2P^2\over 2l_s\Gamma}
\label{eq:7e}
\end{equation}
Till now one can use all the analytical solutions
concerning the random motion of a molecule governed by eq. \ref{eq:2}
by a simple substitution described in eq. \ref{eq:7e}.
Under the condition, $k^2(s,z)>>{\Gamma^2\over 4}$ (adiabatic condition),
one gets \cite{Chand}:
$$<y^2>={kT\over m k^2(s,z)}+\left(y_0^2-{kT\over m k^2(s,z)}\right)
\left(\cos(k_1s)+{\Gamma\over 2k_1}\sin(k_1s)\right)^2\exp(-\Gamma s)$$
$$={\sigma_y^2l_s\over 2\gamma(s,z)\gamma(0)Gk^2(s,z)}
\left({e^2N_eW_{\perp}(z)\over m_0c^2}\right)^2+$$
\begin{equation}
\left(y_0^2-
{\sigma_y^2l_s\over 2\gamma(s,z)\gamma(0)Gk^2(s,z)}
\left({e^2N_eW_{\perp}(z)\over m_0c^2}\right)^2\right)
\left(\cos(k_1s)+{\Gamma\over 2k_1}\sin(k_1s)\right)^2\exp(-\Gamma s)
\label{eq:7x}
\end{equation}
$$<y'^2>={kT\over m }+{k(s,z)\over k_1^2}\left(y_0^2-{kT\over m k^2(s,z)}\right)
\sin^2(k_1s)\exp(-\Gamma s)$$
$$={\sigma_y^2l_s\over 2\gamma(s,z)\gamma(0)Gk^2(s,z)}
\left({e^2N_eW_{\perp}(z)\over m_0c^2}\right)^2+$$
\begin{equation}
{k(s,z)\over k_1^2}\left(y_0^2-
{\sigma_y^2l_s\over 2\gamma(s,z)\gamma(0)Gk^2(s,z)}
\left({e^2N_eW_{\perp}(z)\over m_0c^2}\right)^2\right)
\sin^2(k_1s)\exp(-\Gamma s)
\label{eq:7y}
\end{equation}
$$<yy'>={k(s,z)^2\over k_1}\left({kT\over mk(s,z)^2 }-y_0^2\right)
\left(\cos (k_1s)+{\Gamma \over 2k_1}\sin(k_1s)\right)\exp(-\Gamma s)$$
$$={k(s,z)^2\over k_1}\left(
{\sigma_y^2l_s\over 2\gamma(s,z)\gamma(0)Gk^2(s,z)}
\left({e^2N_eW_{\perp}(z)\over m_0c^2}\right)^2-y_0^2\right)
\times$$
\begin{equation}
\left(\cos(k_1s)+{\Gamma \over 2k_1}\sin(k_1s)\right)\exp(-\Gamma s)
\label{eq:7z}
\end{equation}
where $k_1=\sqrt{k(s,z)^2-{1\over 4}\Gamma^2}$.
The asymptotical values for $<y^2>$, $<y'^2>$, and $<yy'>$ as $s \rightarrow
\infty$ are approximately expressed as: 
\begin{equation}
<y^2>={kT\over m k^2(s,z)}
={\sigma_y^2l_s\over 2\gamma(s,z)\gamma(0)Gk^2(s,z)}
\left({e^2N_eW_{\perp}(z)\over m_0c^2}\right)^2
\label{eq:7}
\end{equation}
\begin{equation}
<y'^2>=k^2(s,z)<y^2>={\sigma_y^2l_s\over 2\gamma(s,z)\gamma(0)G}
\left({e^2N_eW_{\perp}(z)\over m_0c^2}\right)^2
\label{eq:8}
\end{equation}
\begin{equation}
<yy'>=0
\label{eq:8a}
\end{equation}
Inserting eqs. \ref{eq:7}, \ref{eq:8}, and \ref{eq:8a} into the definitions of
the r.m.s. emittance and the normalized r.m.s. emittance shown in 
eqs. \ref{eq:3} and \ref{eq:3a}:
\begin{equation}
\epsilon_{rms}=\left(<y^2><y'^2>-<yy'>^2\right)^{1/2}
\label{eq:3}
\end{equation}
\begin{equation}
\epsilon_{n,rms}=\gamma (s,z)\left(<y^2><y'^2>-<yy'>^2\right)^{1/2}
\label{eq:3a}
\end{equation}
one gets
\begin{equation}
\epsilon_{rms}={\sigma_y^2l_s\over 2\gamma(s,z)\gamma(0)Gk(s,z)}
\left({e^2N_eW_{\perp}(z)\over m_0c^2}\right)^2
\label{eq:9}
\end{equation}
and 
\begin{equation}
\epsilon_{n,rms}={\sigma_y^2l_s\over 2\gamma(0)Gk(s,z)}
\left({e^2N_eW_{\perp}(z)\over m_0c^2}\right)^2
\label{eq:9a}
\end{equation}
\par
The effects of energy dispersion within the bunch can be discussed through
$\gamma (s,z)$, $k^2(s,z)$, and $W_{\perp}(z)$, 
such as BNS damping \cite{Balakin}. 
From eqs. \ref{eq:7}, \ref{eq:9}, and \ref{eq:9a} 
one finds that there are three 
convenient types of variations of $k(s,z)$ with respect to $s$.
If one takes $k^2(s,z)\gamma (s,z)=k^2(0,z)\gamma (0,z)$, one gets $<y^2>$
independent of $s$. If one takes, however, 
$k(s,z)\gamma (s,z)=k(0,z)\gamma (0,z)$, $\epsilon_{rms}$ 
is independent of $s$,
and finally, if $k(s,z)=k(0,z)$, one has $\epsilon_{n,rms}$ is independent
of $s$. One takes usually the first scaling law in accordance with BNS damping.
To calculate the emittance growth
of the whole bunch one has to make an appropriate average  
over the bunch, say Gaussian as assumed above, as follows:
\begin{equation}
\epsilon^{bunch}_{n,rms}=
{\int_{-\infty}^{\infty}\rho (z')\epsilon_{n,rms} (z')dz'\over 
\int_{-\infty}^{\infty}\rho (z')dz'}
\label{eq:9f}
\end{equation} 
To make a rough estimation one can replace $\rho (z)$ by a delta function
$\delta (z-z_c)$,
and in this case the bunch emittance can be still expressed by eq. \ref{eq:9a}
with $W_{\perp}(z)$ replaced by $W_{\perp}(z_c)$,  
where $z_c$ is the center of the bunch.
\par
\section{Method two: Fokker-Planck equation}
Keeping the physical picture described above in mind, one can start
directly with Fokker-Planck equation which governs the distribution
function of the Markov random variable, $y'$:
\begin{equation}
{\partial F(s,y')\over \partial s}=-{\partial \over \partial y'}
\left(AF(s,y')\right)
+{1\over 2}{\partial^2 \over \partial y'^2}(DF(s,y'))
\label{eq:13}
\end{equation}
with
\begin{equation}
A={<<\Delta y'>>\over l_s}
\label{eq:14}
\end{equation}
\begin{equation}
D={<<(\Delta y')^2>>\over l_s}
\label{eq:15}
\end{equation}
where $\Delta y'$ is the increment of $y'$ over $l_s$, and $<<$ $>>$ denotes
the average over a large number of a given type of 
possible structure misalignment error
distributions (in numerical simulations, this average corresponds to
the average over the results obtained from a large number of different seeds,
for a given type of structure misalignment error distribution function, say,
Gaussian distribution).
From eq. \ref{eq:2} one gets the increment of $y'$ over
$l_s$ :
\begin{equation}
\Delta y'=(1-\exp(-{\Gamma l_s\over 2}))y'+\Gamma l_s
\label{eq:15a}
\end{equation}
In consequence, one obtains:
\begin{equation}
<<(\Delta y')>>\approx (1-\exp(-{\Gamma l_s\over 2}))y'
\label{eq:16}
\end{equation}
\begin{equation}
<<(\Delta y')^2>>\approx (1-\exp(-{\Gamma l_s\over 2}))^2y'^2
+<<(\Lambda l_s)^2>>
\exp(-\Gamma l_s)
\label{eq:17}
\end{equation}
where $<<\Gamma l_s>>=0$ has been used. Inserting eqs. \ref{eq:16}
and \ref{eq:17} into eq. \ref{eq:13}, one gets:
$$l_s{\partial F(s,y')\over \partial s}=-(1-\exp(-{\Gamma l_s\over 2}))
{\partial y'F(s,y')\over \partial y'}$$
\begin{equation}
+{(1-\exp(-{\Gamma l_s\over 2}))^2\over 2}{\partial^2 (y'^2F(s,y'))
\over \partial y'^2}
+{<<(\Lambda l_s)^2>>\over 2}\exp(-\Gamma l_s){\partial^2 F\over \partial y'^2}
\label{eq:18}
\end{equation}
Multiplying both sides with $y'^2$ and integrating over $y'$, one has:
\begin{equation}
l_s{d<y'^2>\over ds}=-(1-\exp(-\Gamma l_s))<y'^2>+<<(\Lambda l_s)^2>>
\exp(-\Gamma l_s)
\label{eq:19}
\end{equation}
Assuming that $\Gamma l_s<< 1$, eq. \ref{eq:19} is reduced to:
\begin{equation}
l_s{d<y'^2>\over ds}=-\Gamma l_s<y'^2>+<<(\Lambda l_s)^2>>
\label{eq:20}
\end{equation}
Solving eq. \ref{eq:20}, one gets:
\begin{equation}
<y'^2>={<<\Lambda^2>>l_s\over 2\Gamma }\left(1-\exp\left({-\Gamma s}
\right)\right)+\exp\left({-\Gamma s}\right)y'^2_0
\label{eq:21}
\end{equation}
where $y'_0$ is the initial condition.
Apparently, when $s \rightarrow \infty$, one has:
\begin{equation}
<y'^2>_{\infty}={<<\Lambda^2>>l_s\over 2\Gamma}
={\sigma_y^2l_s\over 2\gamma(s,z)\gamma(0)G}
\left({e^2N_eW_{\perp}(z)\over m_0c^2}\right)^2
\label{eq:22}
\end{equation}
where $\sigma_y^2=<<{y(s,0)}^2>>$.
Eq. \ref{eq:22} is the same as what we have obtained in
eq. \ref{eq:8}. In fact, by solving directly Fokker-Planck equation we obtain
the same set of asymptotical formulae derived in section 3. 
\par   
\section{Multibunch emittance growth}
\begin{figure}[h]
\vspace{9.0cm}
\vskip 2. true cm
\includegraphics{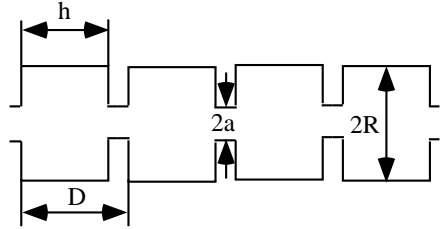}
\vskip -7.5 true cm
\caption{A disk-loaded accelerating structure.  
\label{fig:1}}
\end{figure}

Physically,  the multibunch emittance growth is quite similar to that of the
single bunch case, and each assumed point like bunch in the train can be 
regarded as a slice in
the previously described single bunch. Obviously, 
the slice emittance expressed
in eq. \ref{eq:9a} should be a good starting point for us to estimate the 
emittance growth of the whole bunch train. Before making use of eq. \ref{eq:9a}
let's first look at the differential equation which governs
the transverse motions of the bunch train:
\begin{equation}
{d\over ds}\left(\gamma_n(s){dy_n\over ds}\right)+\gamma_n(s)k_n^2y_n=
{e^2N_e\over m_0c^2}\sum_{i=1}^{n-1}W_{T}\left((n-i){s_b}\right)y_i
\label{eq:26}
\end{equation}
where the subscript $n$ denotes the bunch number, $s_b$ is the distance
between two adjacent bunches, $N_e$ is the particle number in each bunch,
$W_T(s)$ is the long range wakefield produced by each point like bunch
at distance of $s$. 
Clearly, the behaviour of the $ith$ bunch suffers from influences coming from 
all the bunches before it, and we will treat one bunch after another in an
iterative way. First of all, we discuss about the long range wakefields.
Due to the decoherence effect in the long range wakefield only
the first dipole mode will be considered. 
For a constant impedance structure as shown 
in Fig. 1, one has: 
\begin{equation}
W_{T,1}(s)={2ck_1\over \omega_{1}a^2}
\sin(\omega_{1}{s\over c})
\exp\left({-{\omega_{1}\over 2Q_{1}}\left({s\over c}\right)}\right)
\exp\left(-{\omega^2_{1}\sigma_z^2\over 2c^2}\right)
\label{eq:777}
\end{equation}
where $\sigma_z$ is the rms bunch length ($\sigma_z$ is used
to calculate the transverse wake potential, and the point charge assumption is
still valid), $\omega_{1}$ and $Q_{1}$
are the angular frequency and the loaded quality factor of the dipole mode,
respectively. The loss factor $k_1$ in eq. \ref{eq:777} 
can be calculated analytically as \cite{Gao2}:
\begin{equation}
k_1={hJ_1^2\left({u_{11}\over R}a\right)\over \epsilon_0\pi DR^2
J_2^2(u_{11})}S(x_1)^2
\label{eq:10}
\end{equation}
\begin{equation}
S(x)={\sin x \over x}
\label{eq:11}
\end{equation}
\begin{equation}
x_1={hu_{11}\over 2R}
\label{eq:12}
\end{equation}
where $R$ is the
cavity radius, $a$ is the iris radius, $h$ is the cavity hight
as shown in Fig. \ref{fig:1},
and $u_{11}=3.832$ is the first root of the
first order Bessel function.\linebreak


\noindent
 To reduce the long range wakefield
one can detune and damp the concerned dipole mode. The resultant
long range wakefield of the detuned and damped structure (DDS) can be 
expressed as:
\begin{equation}
W_{T,DDS}(s)={1\over N_c}
\sum_{i=1}^{N_c}{2ck_{1,i}\over \omega_{1,i}a_i^2}
\sin(\omega_{1,i}{s\over c})
\exp\left({-{\omega_{1,i}\over 2Q_{1,i}}\left({s\over c}\right)}\right)
\exp\left(-{\omega^2_{1,i}\sigma_z^2\over 2c^2}\right)
\label{eq:7777}
\end{equation}
where $N_c$ is the number of the cavities in the structure.
When $N_c$ is very large one can use following formulae to describe ideal
uniform and Gaussian detuning structures \cite{Gao3}:
\hfill \break
1) Uniform detuning:
\begin{equation}
W_{T,1,U}=2<K>\sin\left({2\pi <f_1>s\over c}\right)
{\sin(\pi s\Delta f_1/c)\over (\pi s\Delta f_1/c)}
\exp\left(-{\pi <f_1>s\over <Q>_1c}\right)
\label{eq:17}
\end{equation}
\hfill \break
2) Gaussian detuning:
\begin{equation}
W_{T,1,G}=2<K>\sin\left({2\pi <f>s\over c}\right)e^{-2(\pi \sigma_fs/c)^2}
\exp\left(-{\pi <f>s\over <Q>_1c}\right)
\label{eq:18}
\end{equation}
where $K={ck_{1,i}\over \omega_{1,i}a_i^2}$, $f_1={\omega_1\over 2\pi}$,
$\Delta f_1$ is full range the synchronous 
frequency spread due to the detuning effect,
$\sigma_f$ is the $rms$ width in Gaussian frequency distribution.
\par
Once the long range wakefield is known one can use eq. \ref{eq:7} to estimate
$<y_i^2>$ in an iterative way, and the emittance
of the whole bunch can be calculated accordingly as we will show later. 
For example, if a bunch train is injected on axis 
($y_n=0$) into the main linac
of a linear collider with structure {\it rms} misalignment $\sigma_y$, 
at the end of the linac one has:
\begin{equation}
<y_1^2>=0
\label{eq:200}
\end{equation} 
\begin{equation}
<y_2^2>={(\sqrt{(\sigma_y^2+{<y_1^2>\over 2})}e^2N_e\vert W_{T}(s_b)\vert  )^2(s_b)l_s\over 2\gamma(s)\gamma(0)Gk_n^2(s)( m_0c^2)^2}
\label{eq:201}
\end{equation} 
\begin{equation}
<y_3^2>={\left(\sqrt{(\sigma_y^2+{<y_1^2>\over 2})}e^2N_e\vert W_{T}(2s_b)\vert
+\sqrt{(\sigma_y^2+{<y_2^2>\over 2})}e^2N_e\vert W_{T}(s_b)\vert \right)^2
l_s\over 2\gamma(s)\gamma(0)Gk_n^2(s)( m_0c^2)^2}
\label{eq:202}
\end{equation} 
and in a general way, one has:
\begin{equation}
<y_i^2>={\left(\sum_{j=1}^{i-1}\sqrt{(\sigma_y^2+{1\over 2}<y_j^2>)}
e^2N_e\vert W_{T}((i-j)s_b)\vert\right)^2
l_s\over 2\gamma(s)\gamma(0)Gk_n^2(s)( m_0c^2)^2}
\label{eq:203}
\end{equation} 
Finally, one can use the following formula to estimate the projected emittance 
of the bunch train:
\begin{equation}
\epsilon_{n,rms}^{train}={\gamma (s)\overline{k(s)}\over N_b}\sum_{i=1}^{N_b}<y_i^2>
\label{eq:205}
\end{equation}
where $k(s)=k_n(s)$ (since the bunch to bunch energy spread has been ignored),
and $\overline{k(s)}$ is the average over the linac.
\par
It is high time now for us to point out that the analytical expressions
for the single and multibunch emittance growths established above give the 
statistical results of infinite number of machines with Gaussian structure
misalignment error distribution, which corresponds to using infinite seeds
in numerical simulations.
\par
\section{Comparison with numerical simulation results}
\begin{figure}[h]
\vspace{8.0cm}
\vskip 1. true cm
\includegraphics{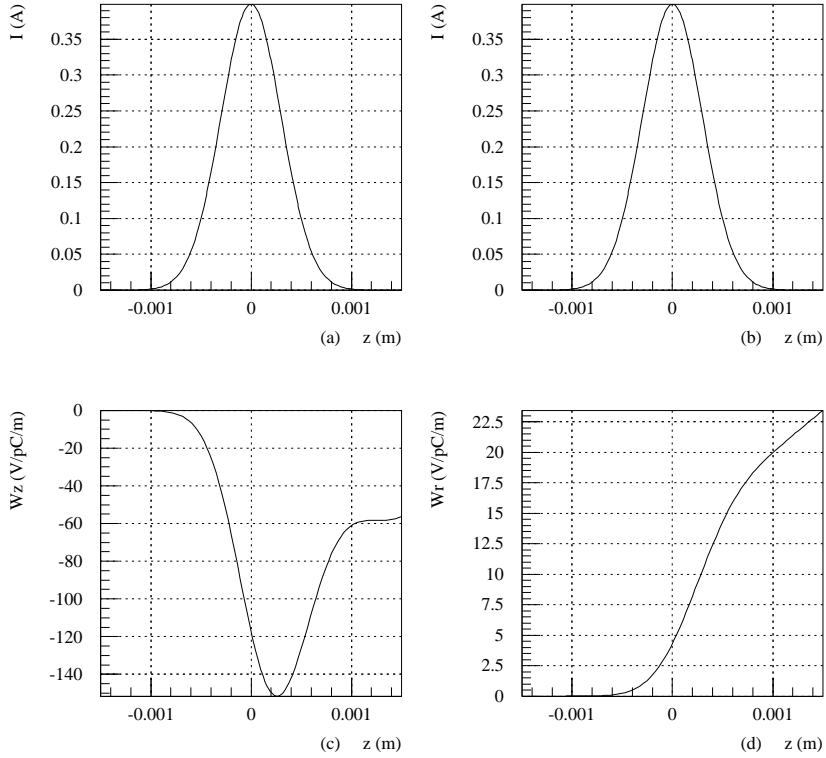}
\vskip 2. true cm
\caption{The short range wakefields of SBLC type structure with
$\sigma_z=300$ $\mu$m, and the beam iris $a=0.0133$ m. (a) and (b) are
the bunch current distributions. (c) is monopole the longitudinal wakefield.
(d) is the dipole transverse wakefield at $r=a$.  }
\label{fig1}
\end{figure}

To start with, we take the single bunch emittance growth in the
main linac of SBLC \cite{SBLC} for example. 
The short range wakefields in the accelerating 
S-band structures are obtained by using the analytical formulae \cite{Gao2}
and shown in Fig. \ref{fig1}. In the main linac the beam
is injected at 3 GeV and accelerated to 250 GeV with an accelerating
gradient of 17 MV/m. The accelerating structure
length $l_s=$6 m, the average beta function $\overline{\beta (s)}$ is about 70 m
($k(s,z)={1\over \beta(s)}$ for smooth focusing), 
the bunch population $N_e=1.1\times 10^{10}$,
the bunch length $\sigma_z=300$ $\mu$m, and 
and corresponding dipole mode short range wakefield 
$W_{\perp}(z_c)=338$ V/pC/m$^2$.
Inserting these parameters into eq. \ref{eq:9a}, one finds 
$\epsilon_{n,rms}=8.66\times \sigma_z^2$. If accelerating structure misalignment
error $\sigma_y=100$ $\mu$m, one gets a normalized emittance growth of 
8.66$\times 10^{-8}$ mrad, i.e., 35$\%$ increase 
compared with the designed normalized emittance of $2.5\times 10^{-7}$ mrad.
The analytical result agrees quite well with that obtained from numerical 
simulations \cite{SBLC}.  Now, we apply the analytical formulae established for the multibunch emittance
growth
to SBLC, TESLA and NLC linear collider projects where enormous numerical simulations have been done. The machine parameters are given in Tables 1 to 4
which have been used in the analytical calculation in this paper.
Firstly, we look at SBLC.


\newpage

\begin{figure}[h]
\vspace*{8.cm}
\vspace*{3mm}
\includegraphics{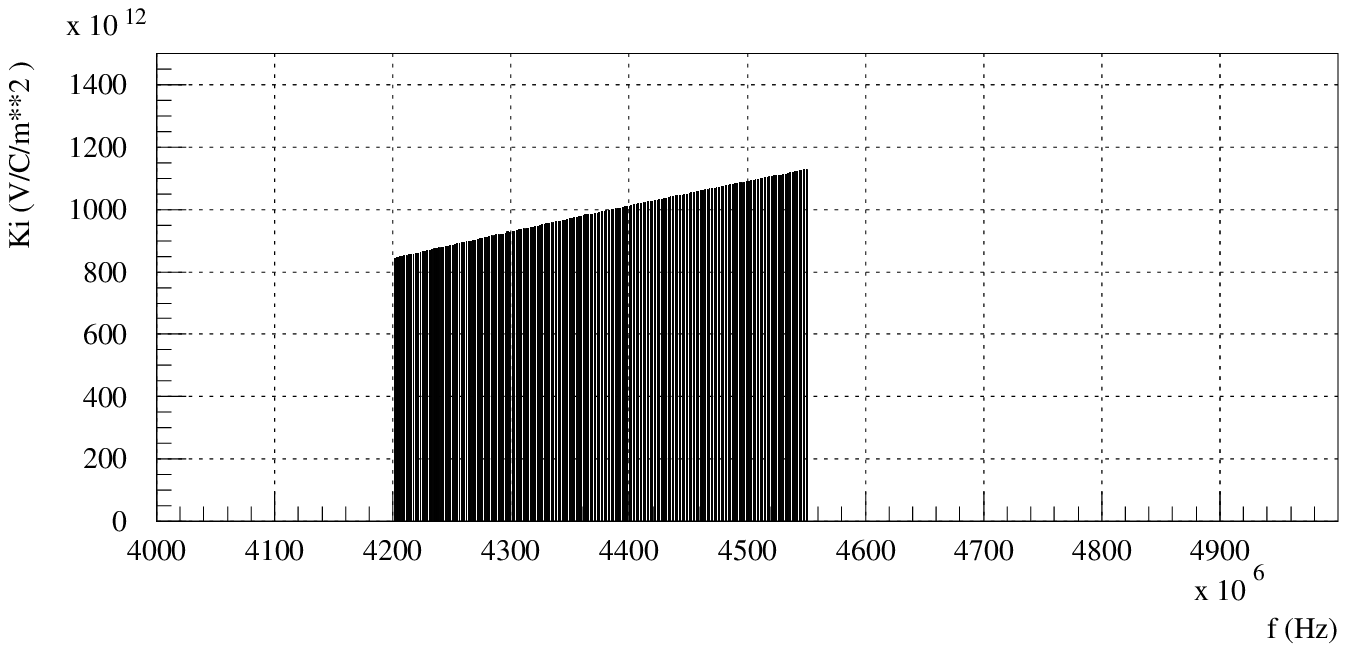}
\vskip -3.8 true cm
\caption{The $K_i$ vs dipole mode frequency (SBLC).  
\label{fig4}}
\end{figure}

\noindent
 Fig. \ref{fig4} shows the ``kick factor'' $K$ defined in eqs. \ref{eq:17} and \ref{eq:18} vs the dipole mode frequency.\\

\begin{figure}[h]
\vspace*{7.cm}
\vskip 1. true cm
\includegraphics{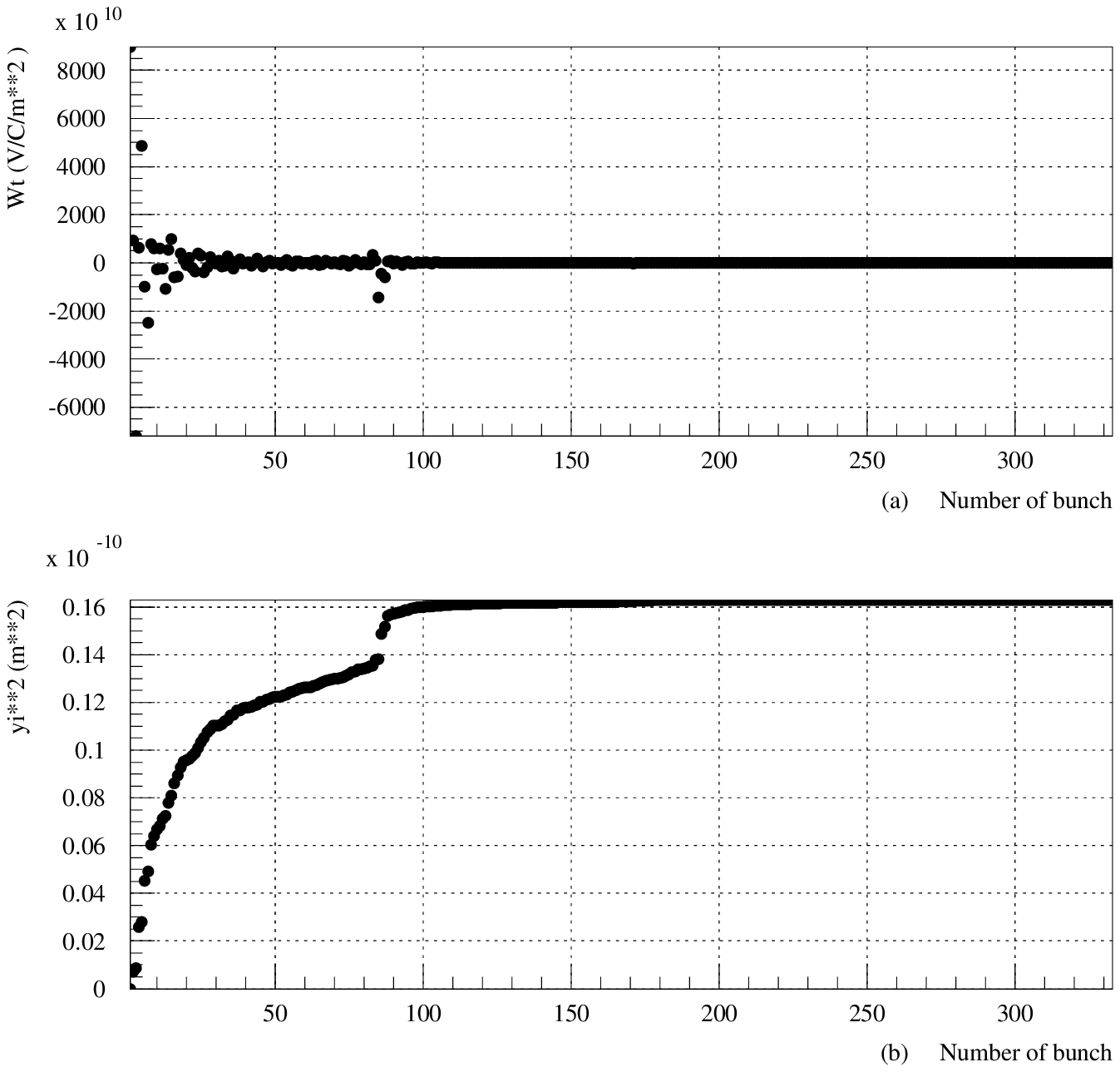}
\vskip 1.5 true cm
\caption{(a) the long range dipole mode wakefield vs the number of bunch.
(b) the $y_i^2$ at the end of linac vs the number of bunch 
(SBLC, $Q_1=2000$, $\sigma_y=100 \ \mu$m).}
\label{fig5}
\end{figure}

\begin{figure}[h]
\vspace*{4.0cm}
\vskip 0.5 true cm
\includegraphics{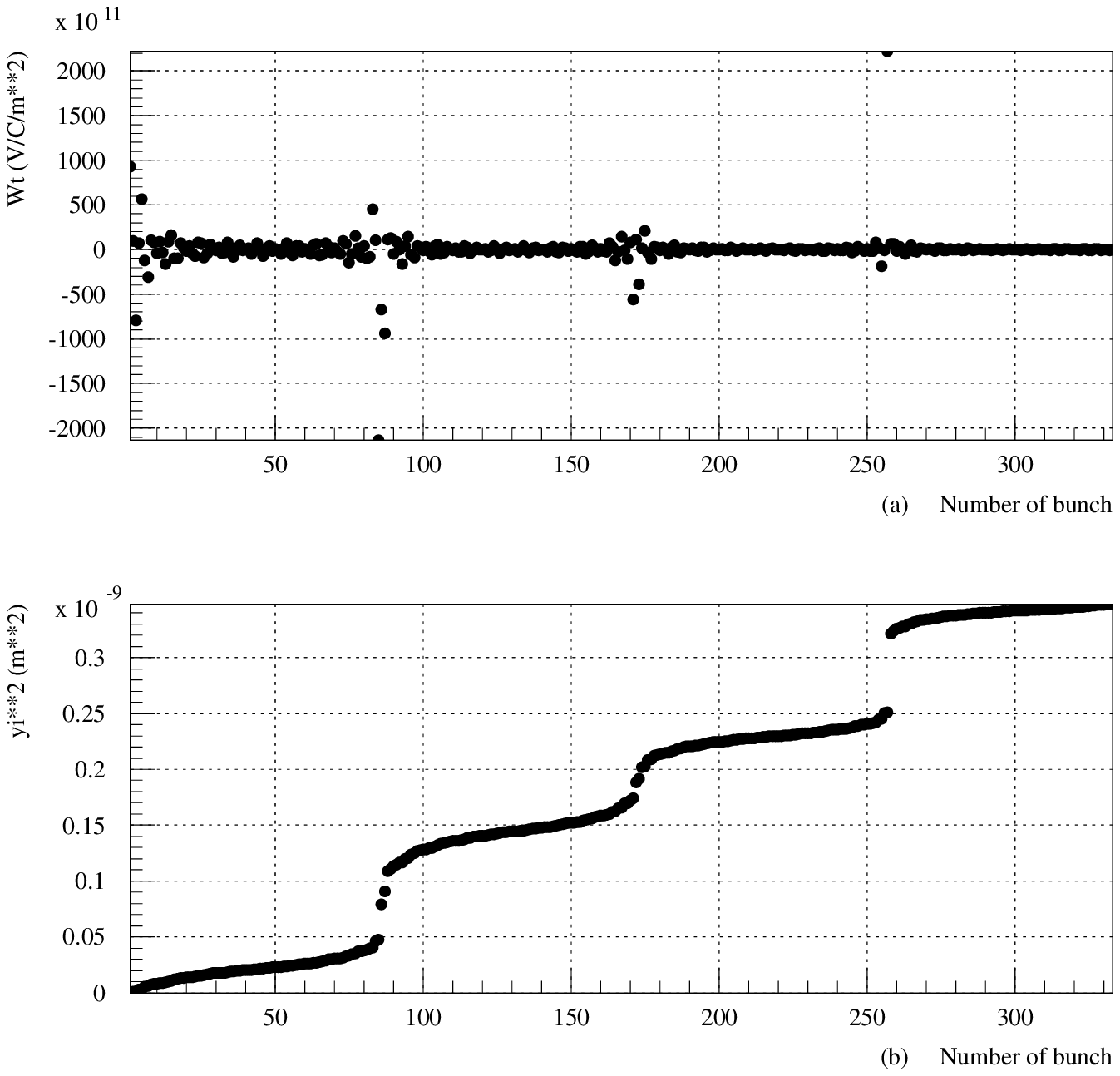}
\vspace*{1.4cm}
\caption{(a) the long range dipole mode wakefield vs the number of bunch.
(b) the $y_i^2$ at the end of linac vs the number of bunch 
(SBLC, $Q_1=10000$, $\sigma_y=100 \ \mu$m).}
\label{fig6}
\end{figure}

\noindent
 Fig. \ref{fig5}(a) gives the long range transverse wakefield produced by the first bunch at the locations where find the succeeding bunches, while 
Fig. \ref{fig5}(b) illustrates the square of the {\it rms} deviation of each
bunch at the end of the linac with the dipole loaded quality factor 
$Q_1=2000$. The corresponding results for $Q_1=10000$ are  shown in 
Fig. \ref{fig6}. The normalized emittance growths compared with the 
design value at the interaction point ($\epsilon_{n,rms}^{design,IP}=2.5
\times 10^{-7}$ mrad) are 32$\%$ and 388$\%$ corresponding to the two cases,
respectively as shown in Table 4, which agree well with the numerical results \cite{SBLC}.
\newpage

\quad
\vspace{40mm}

\begin{figure}[h]
\vspace{8.0cm}
\vskip 1. true cm
\includegraphics{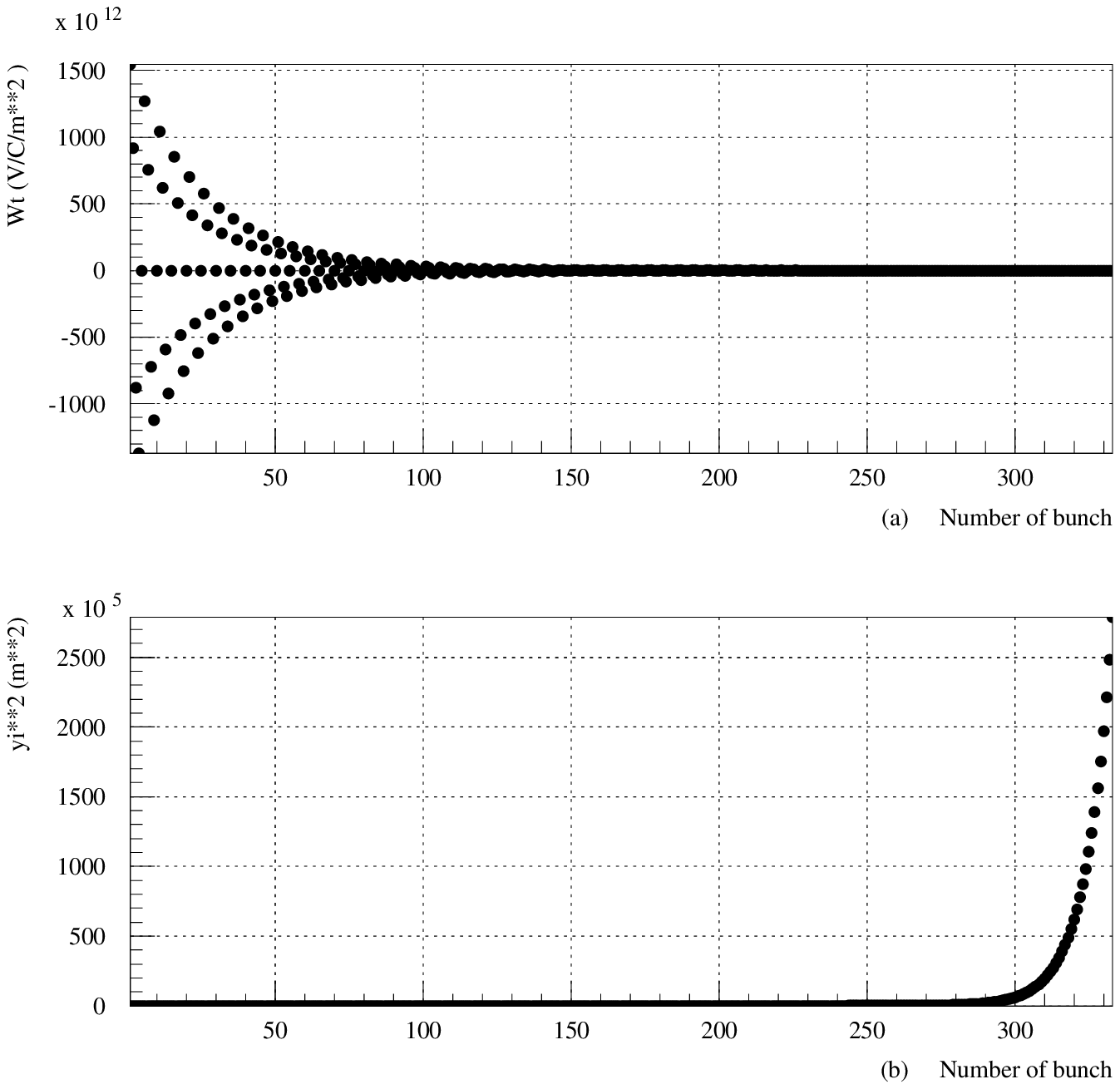}
\vskip 2. true cm
\caption{(a) the long range dipole mode wakefield vs the number of bunch.
(b) the $y_i^2$ at the end of linac vs the number of bunch (SBLC no detuning,
$Q_1=2000$, $\sigma_y=100 \ \mu$m).
\label{fig5a}}
\end{figure}

\begin{figure}[h]
\vspace{7.9cm}
\vskip 1. true cm
\includegraphics{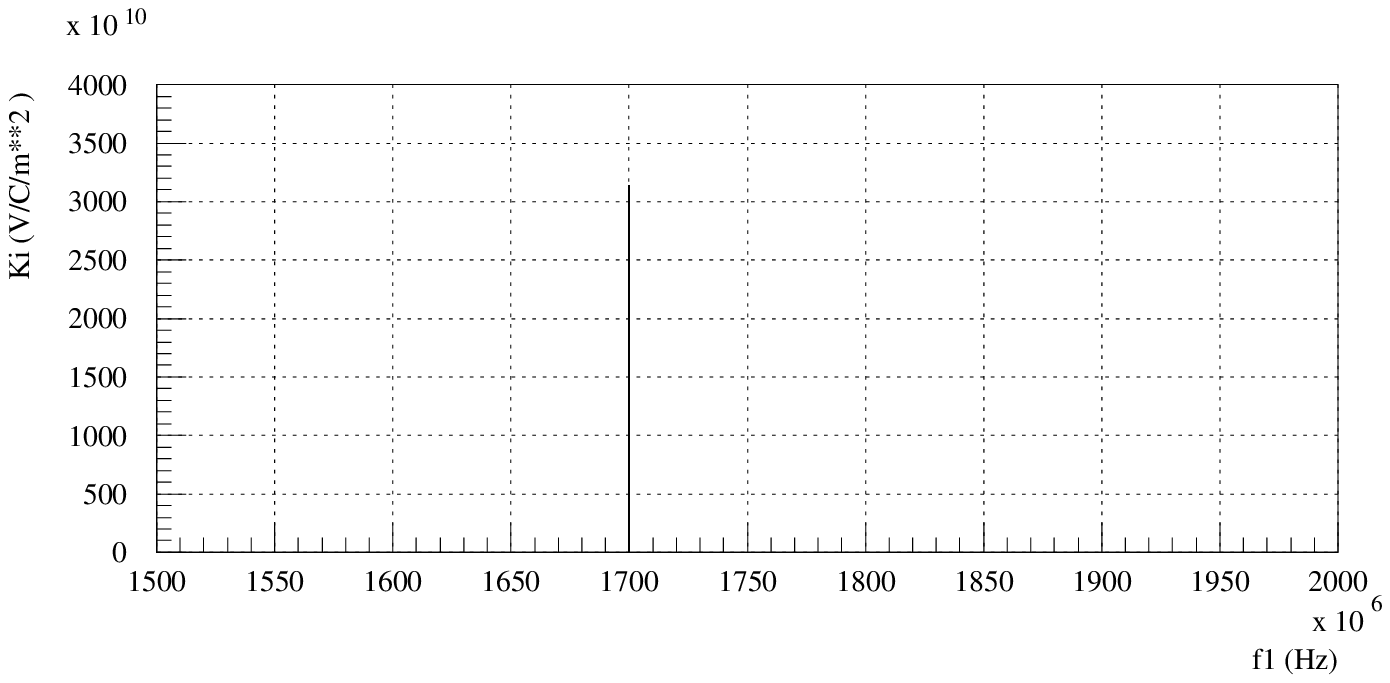}
\vspace{-40 mm} 
\caption{The $K_i$ vs dipole mode frequency (TESLA). 
\label{fig7}}
\end{figure}

\newpage
\noindent
To demonstrate the necessity of detuning cavities we show the violent 
bunch train blow up if constant impedance structures are used in spite of
$Q_1$ being loaded to 2000 as shown in Fig. \ref{fig5a}. Secondly, TESLA (the version appeared in ref. 10) is investigated. From Fig. \ref{fig7} one agrees that it is a no detuning case.


\begin{figure}[h]
\vspace{7.8cm}
\vskip 1.3 true cm
\includegraphics{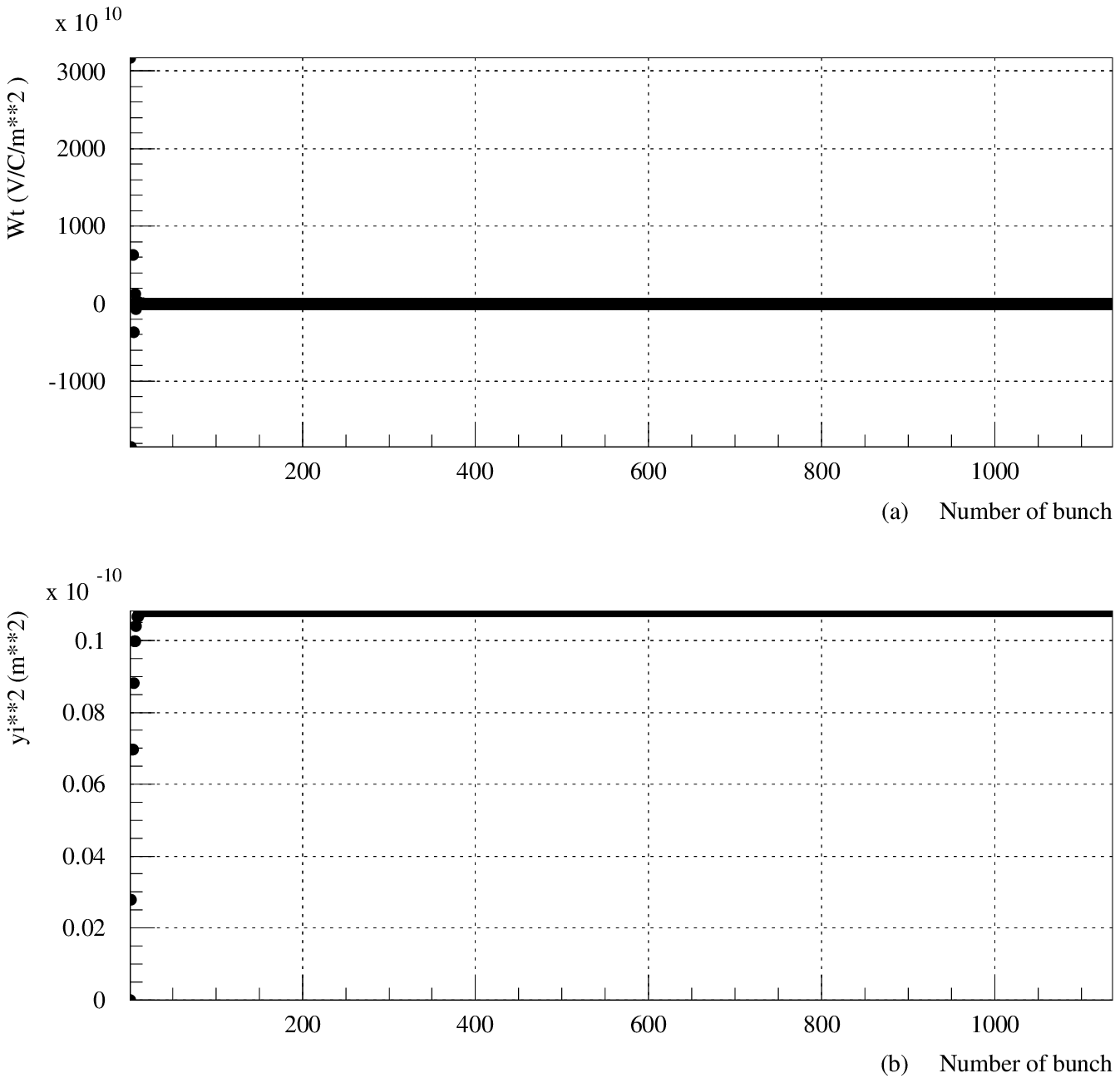}
\vskip 1.5 true cm
\caption{(a) the long range dipole mode wakefield vs the number of bunch.
(b) the $y_i^2$ at the end of linac vs the number of bunch 
(TESLA, Q1=7000, $\sigma_y=500 \ \mu$m).
\label{fig8}}
\end{figure}

\begin{figure}[h]
\vspace{8.0cm}
\vskip 1. true cm
\includegraphics{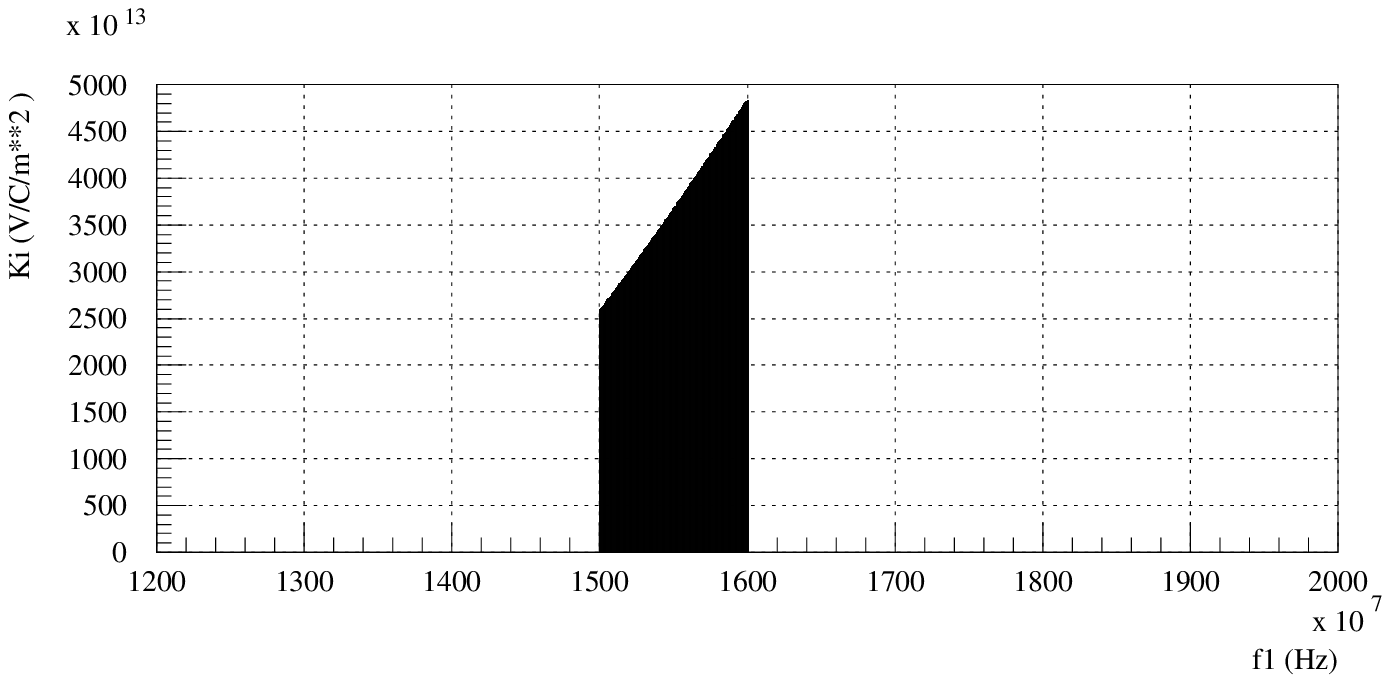}
\vskip -4. true cm
\caption{The $K_i$ vs dipole mode frequency (NLC X-band linac). 
\label{fig9}}
\end{figure}

\begin{figure}[h]
\vspace{8.0cm}
\vskip 1. true cm
\includegraphics{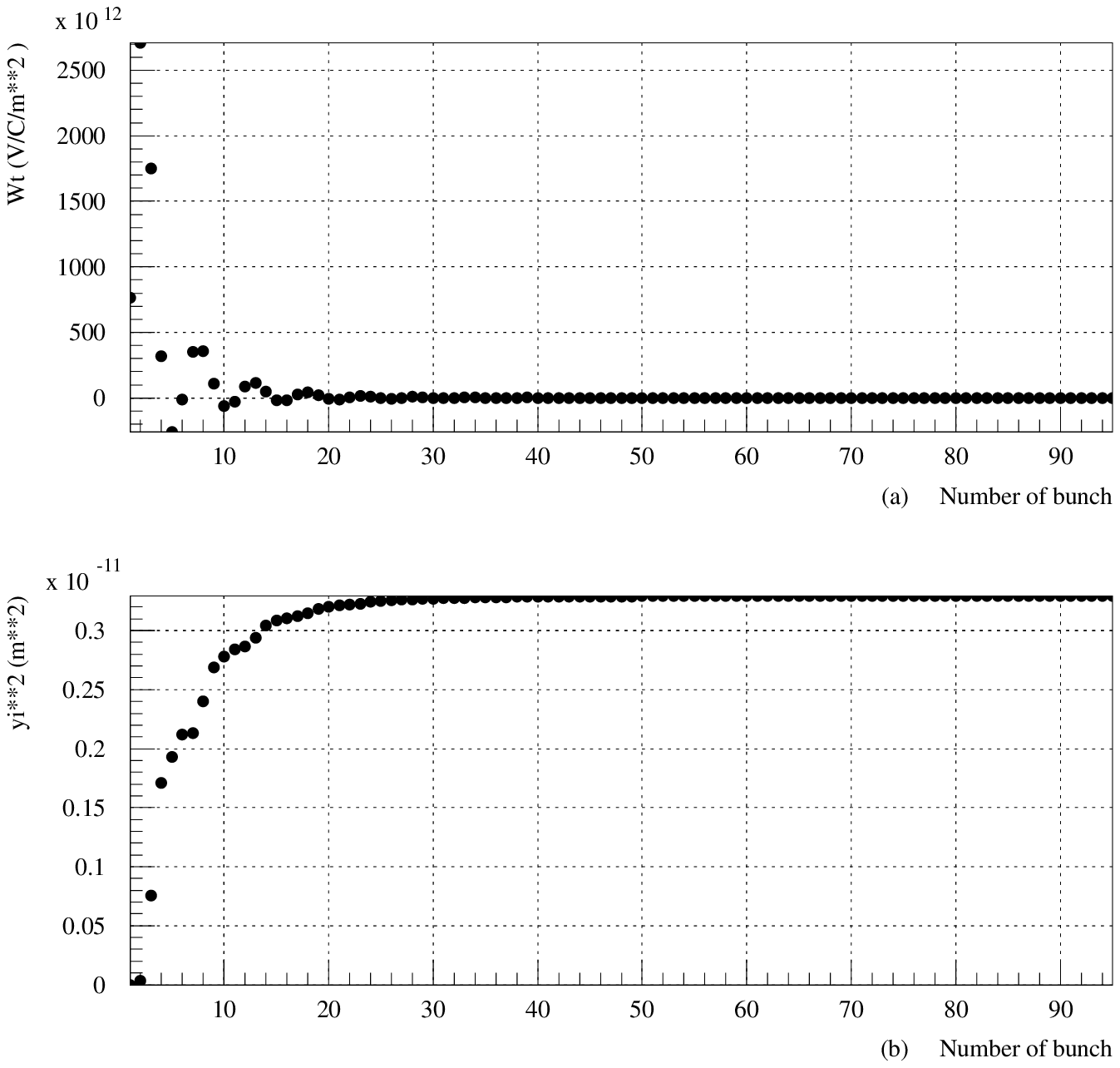}
\vskip 2. true cm
\caption{(a) the long range dipole mode wakefield vs the number of bunch.
(b) the $y_i^2$ at the end of linac vs the number of bunch 
(NLC X-band linac, $Q_1$=1000, $\sigma_y=15 \ \mu$m).
\label{fig10}}
\end{figure}

\begin{figure}[h]
\vspace{8.0cm}
\vskip 1. true cm
\includegraphics{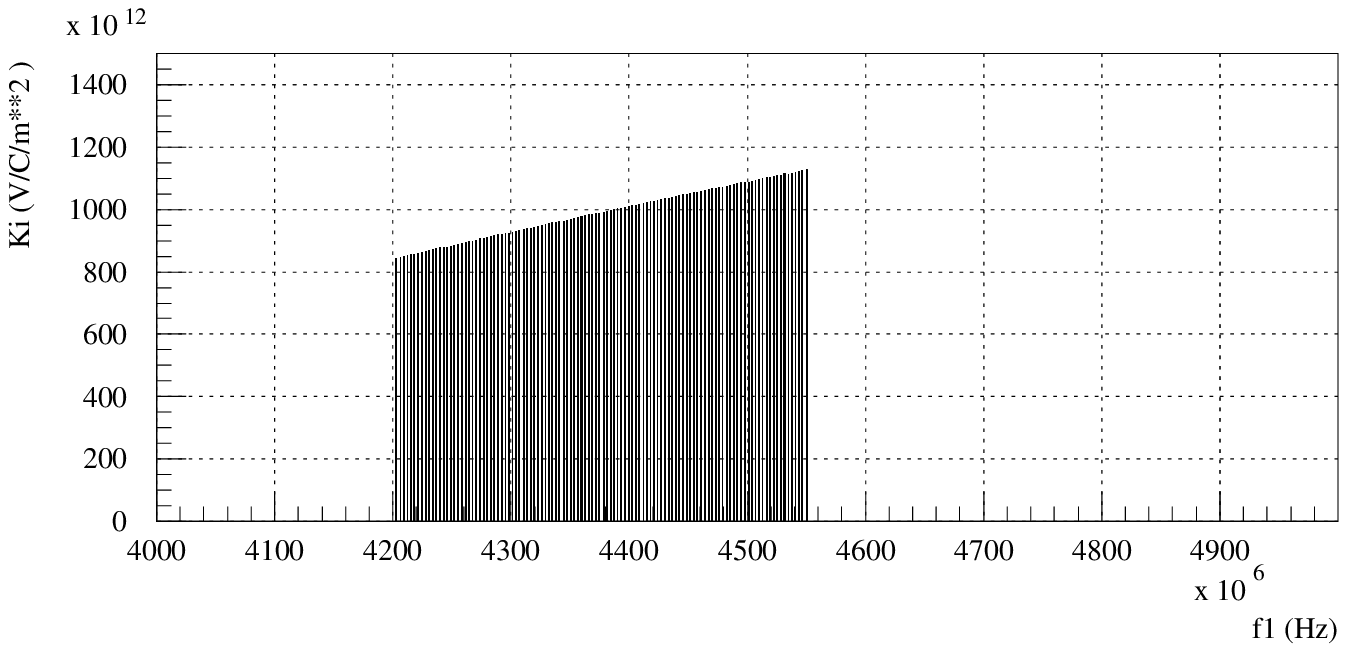}
\vskip -4. true cm
\caption{The $K_i$ vs dipole mode frequency (NLC S-band prelinac). 
\label{fig11}}
\end{figure}

\begin{figure}[h]
\vspace{4.0cm}
\vskip 1. true cm
\includegraphics{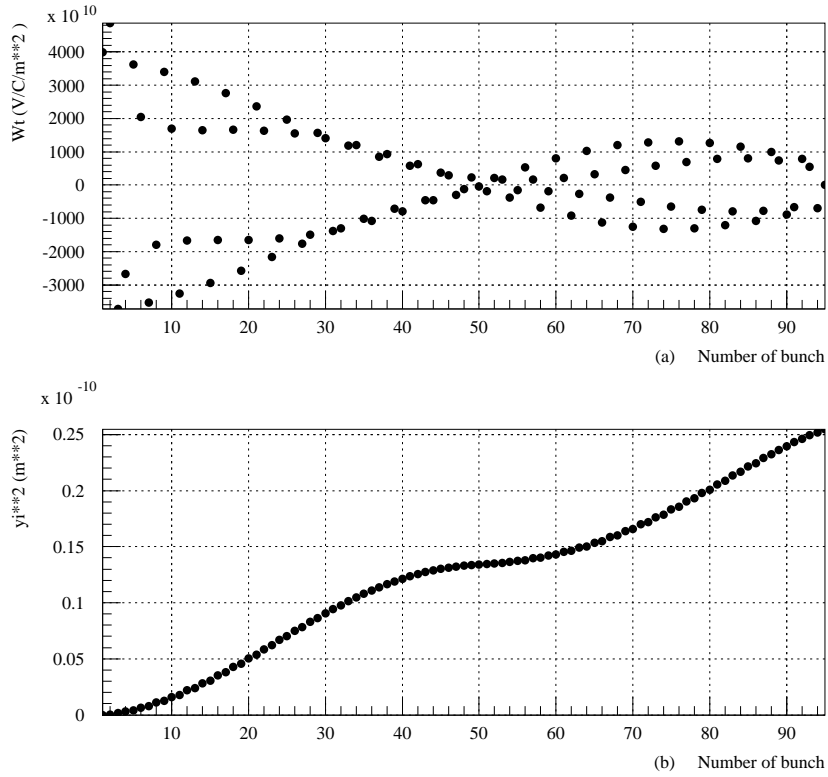}
\vskip 2. true cm
\caption{(a) the long range dipole mode wakefield vs the number of bunch.
(b) the $y_i^2$ at the end of linac vs the number of bunch 
(NLC S-band prelinac, $Q_1$=10000, $\sigma_y=50 \  \mu$m).}
\label{fig12}
\end{figure}

\begin{figure}[h]
\vspace{6.0cm}
\vskip 1. true cm
\includegraphics{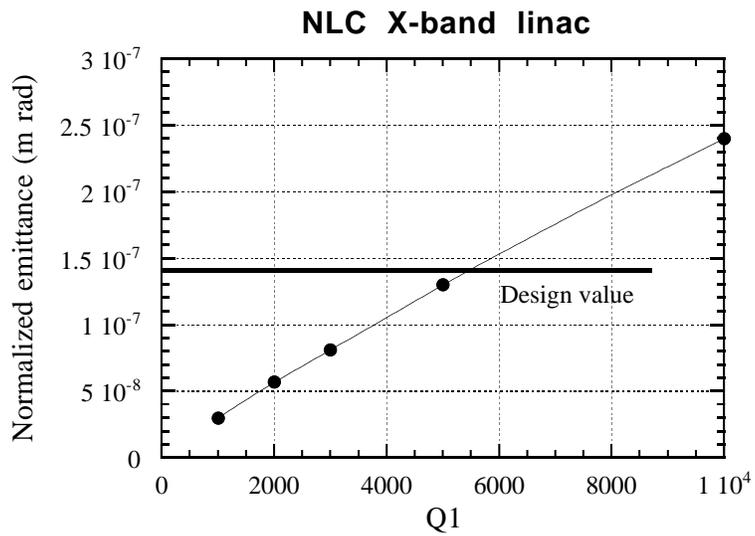}
\vskip 0.1 true cm
\caption{The normalized emittance growth vs $Q_1$ 
with $\sigma_y=15 \ \mu$m (NLC X-band linac). 
\label{fig13}}  
\end{figure}

\begin{figure}[h]
\vspace{7.0cm}
\vskip 1. true cm
\includegraphics{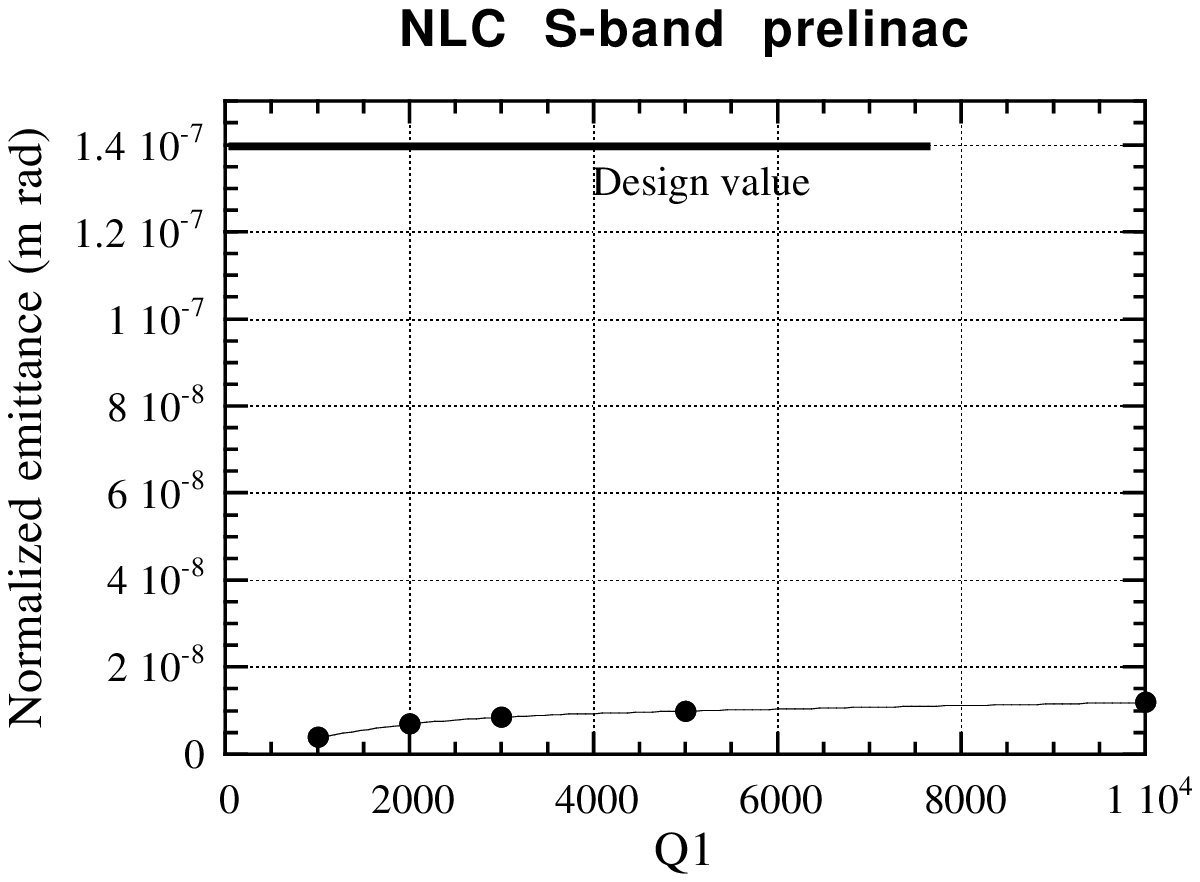}
\vskip 0. true cm
\caption{The normalized emittance growth vs $Q_1$ with $\sigma_y=50 \mu$m (NLC S-band linac). 
\label{fig14}} 
\end{figure}

\quad
\vspace{6mm}
\newpage
From the results shown in Fig. \ref{fig8} and Table 4 
one finds that taking structure misalignment error $\sigma_y=500 \ \mu$m
and $Q_1=7000$ one gets an normalized emittance growth of 24$\%$ which
is a very reasonable result compared what has been found numerically in
ref. 10.
Thirdly, we look at NLC X-band main linac. To facilitate the exercise
we assume the detuning is effectuated as shown in Fig. \ref{fig9}   
(in reality, NLC uses Gaussian detuning). Fig. \ref{fig10} shows the analytical
results with $\sigma_y=15 \ \mu$m and $Q_1=1000$. From Table 4 one finds
a normalized emittance growth of 21$\%$.
Then, we examine NLC S-band prelinac.
Assuming that the detuning of the dipole mode is shown in Fig. \ref{fig11}, 
one gets the multibunch transverse behaviour
and the normalized emittance growth in Fig. \ref{fig12} and Table 4.
Finally, in Figs. \ref{fig13} and \ref{fig14} we give more information about the emittance growth vs $Q_1$ in NLC X-band and S-band linacs.


\quad
\vspace{15mm}
\newpage
\quad
\vspace{4.0cm}

\vspace*{10mm}
\newpage

\begin{table}[h]
\begin{center}
\begin{tabular}{|l|l|l|l|l|l|l|l|}
\hline
Machine&$l_s$ (m)&$N_c$&$f_1$ (GHz)&$a$ (m)&$D$ (m)&h (m)&R (m) \\
\hline
SBLC&6&180&4.2-4.55&0.015-0.01&0.035&0.0292&0.041\\
\hline
TESLA&1&9&1.7&0.035&0.115&0.0974&0.095\\
\hline
NLC X-band&1.8&206&15-16&0.0059-0.00414&0.00875&0.0073&0.011\\
\hline
NLC S-band&4&114&4.2-4.55&0.015-0.01&0.035&0.0292&0.041\\
\hline
\end{tabular}
\end{center}
\caption{The machine parameters I.}
\vspace{-5mm}
\end{table}
\begin{table}[h]
\begin{center}
\begin{tabular}{|l|l|l|l|l|l|l|}
\hline
Machine&$N_e$ ($\times 10^{10}$)&$s_b$ (m)&$E_z$ (MV/m)&
$\sigma_z (\mu$m)&$N_b$&$Q_1$\\
\hline
SBLC&1.1&1.8&17&300&333&2000,10000\\
\hline
TESLA&3.63&212&25&700&1136&7000\\
\hline
NLC X-band&1.1&0.84&50&145&95&1000\\
\hline
NLC S-band&1.1&0.84&17&500&95&10000\\
\hline
\end{tabular}
\end{center}
\caption{The machine parameters II.}
\vspace{-10mm}
\end{table}
\vspace*{-20mm}
\begin{table}[h]
\begin{center}
\begin{tabular}{|l|l|l|l|l|l|}
\hline
Machine&$\gamma (0)$ (GeV/MeV)&$\gamma $ (GeV/MeV)&$\overline{k(s)}$ (1/m)&
$\sigma_y$ ($\mu$m)\\
\hline
SBLC&3/0.511&250/0.511&1/90&100\\
\hline
TESLA&3/0.511&250/0.511&1/90&500\\
\hline
NLC X-band&10/0.511&250/0.511&1/50&15\\
\hline
NLC S-band&20/0.511&10/0.5111&1/20&50\\
\hline
\end{tabular}
\end{center}
\caption{The machine parameters III.}
\vspace{-5mm}
\end{table}
\begin{table}[h]
\begin{center}
\begin{tabular}{|l|l|l|l|}
\hline
Machine&$\epsilon^{train,numeri.}_{n,rms}$ (mrad)&$\epsilon^{train,analy.}_{n,rms}$ (mrad)&$\epsilon^{IP,design}_{n,rms}$ (mrad)\\
\hline
SBLC&2.3$\times 10^{-8},8.8\times 10^{-7}$
&8.$\times 10^{-8},9.7\times 10^{-7}$&2.5$\times 10^{-7}$\\
\hline
TESLA&$\sim$2.5$\times 10^{-8}$&5.9$\times 10^{-8}$&2.5$\times 10^{-7}$\\
\hline
NLC X-band&-&3$\times 10^{-8}$&1.4$\times 10^{-7}$\\
\hline
NLC S-band&-&1.2$\times 10^{-8}$&1.4$\times 10^{-7}$\\
\hline
\end{tabular}
\end{center}
\caption{The normalized train emittance growth.}
\end{table}
\newpage
\section{Conclusion}
We treat the single and multibunch emittance growths in the main linac
of a linear collider in analogy to the 
Brownian motion of a molecule, and obtained the analytical
expressions for the emittance growth due to accelerating structure misalignment errors by solving Langevin equation. 
As proved in this paper, 
the same set of formulae can be derived also
by solving directly Fokker-Planck equation. 
Analytical results have been compared with those coming
from the numerical simulations, such as SBLC and TESLA, and the agreement is 
quite well. As interesting applications, we give the analytical results on
the estimation of the multibunch emittance growth in NLC X-band and S-band
linacs.
\par
\section{Acknowledgement}
It is a pleasure to discuss with J. Le Duff on the stochastic motions. 
I thank F. Richard for his reminding me the work of J. Perez Y Jorba
on inverse multiple Touschek effect in linear colliders,
and T.O. Raubenheimer
for the discussion on the NLC parameters and detailed beam dynamics problems.

\end{document}